\documentclass[doublecol]{epl2}
\usepackage{amsmath}
\usepackage{amssymb}
\usepackage{graphicx}
\usepackage{dcolumn}
\usepackage{bm}
\usepackage{color}
\definecolor{blue}{rgb}{0.3,0.3,0.9}

\title{Emergent BCS regime of the two-dimensional fermionic Hubbard model:
 ground-state phase diagram}

\author{Youjin Deng\inst{1,2}\footnote{yjdeng@ustc.edu.cn} \and Evgeny Kozik\inst{3,4}\footnote{evgeny.kozik@kcl.ac.uk} \and Nikolay V. Prokof'ev\inst{2,5} \and Boris V. Svistunov\inst{1,2,5} }
\institute{
  \inst{1}  Hefei National Laboratory for Physical Sciences at Microscale, Department of Modern Physics, and
              Synergetic Innovation Center of Quantum Information and Quantum Physics,
              University of Science and Technology of China-Hefei, Anhui 230026, China\\
  \inst{2} Department of Physics, University of Massachusetts-Amherst, MA 01003, USA \\
  \inst{3} Department of Physics, King's College London, Strand-London WC2R 2LS, UK \\
  \inst{4} Centre de Physique Th\'eorique, Ecole Polytechnique, CNRS-91128 Palaiseau Cedex, France \\
  \inst{5}  Russian Research Center ``Kurchatov Institute,'' 123182 Moscow, Russia \\
}

\abstract{
For over half a century, the Hubbard model has played a paradigmatic role in attempts to understand quantum phenomena exhibited by correlated electrons in solids. Despite substantial effort and apparent simplicity of the model, its behavior in many important regimes has remained unknown. Here we study superfluidity in the two-dimensional Hubbard model with controlled error bars up to the coupling strength $U=4$ and filling factor $n=0.7$. We show, by means of unbiased diagrammatic Monte Carlo simulations, that in this regime the superfluid transition is governed by Fermi liquid physics with an emergent weak BCS-type coupling driving the instability. The corresponding ground-state phase diagram in the $(n,U)$ plane describes competition between the superfluid states of $p-$ and $d-$wave symmetry. We also report dimensionless coupling constants in this effective BCS regime.
}

\begin{document}
\maketitle
{\it The fermionic Hubbard model}~\cite{Hubbard63, Anderson},
\begin{equation}
\hat{H} = - \sum_{\langle i,j\rangle, \sigma}    \hat{c}_{i \sigma}^{\dagger} \hat{c}_{j \sigma}^{\phantom{\dagger}}
  + U \sum_i  \hat{n}_{i \uparrow} \hat{n}_{i \downarrow} - \mu \sum_{i, \sigma} \hat{n}_{i \sigma}
\label{FH}
\end{equation}
($\hat{c}_{i \sigma}^{\dagger}$ creates a fermion with spin projection $\sigma=\uparrow,  \downarrow$ on site $i$;
$ \hat{n}_{i \sigma} = \hat{c}_{i \sigma}^{\dagger}\hat{c}_{i \sigma}^{\phantom{\dagger}}$;
$\langle \dots \rangle$ restricts summation to neighboring lattice sites;
$U$ and $\mu$ are, respectively, the on-site repulsion and the chemical potential in units of the hopping amplitude) is one of  ``standard models" of condensed matter physics.
The metal-insulator transition at half filling $\langle \hat{n}_{i \uparrow}+ \hat{n}_{i \downarrow} \rangle =1 $, along with  the antiferromagnetism promoted by it, was the main context
of the original formulation of (\ref{FH})
and subsequent two decades of its intensive theoretical studies. The advent of high-temperature superconductivity dramatically enhanced (and changed the focus of)
the interest to Eq.~(\ref{FH}). It became paradigmatic  for high-temperature superconductors \cite{Anderson}, {\it at least} as a minimalistic
Hamiltonian featuring (not far from half filling) the relevant $d_{x^2-y^2}$ Cooper instability, solely due to repulsive interaction between fermions. 
The most recent wave
of interest to Eq.~(\ref{FH}) has been generated by its direct realization with ultracold atoms in optical lattices \cite{Esslinger05,Jordens08, Schneider08}.

Decades of theoretical studies of Cooper instability in the model (\ref{FH})
have seen a number of remarkable successes. Controlled results were obtained
in certain limiting cases:
vanishingly small interaction or/and low filling \cite{MKagan,Baranov,Chubukov92,Chubukov,Zanchi,Halboth,Fukazawa,Hlubina}
and close to half-filling, by a combination of numerical methods including
determinant Monte Carlo \cite{Blankenbecler, Staudt2000, Burovski},
density matrix renormalization group \cite{White}, and the dynamical
mean-field theory on large clusters
\cite{Metzner,Georges1992,Georges1996,Maier,Toschi,Rubtsov,Yang,Chen2012,Chen2013,Gull}.
For the 2D case we are interested in here, it has been found that,
at a fixed filling factor and $U \to 0$  (within the second-order perturbation theory in $U$)
the ground state of the system is either $d_{xy}$-wave (smaller fillings) or  $d_{x^2-y^2}$-wave (higher fillings) BCS superfluid, with a pocket of a $p$-wave phase near quarter filling \cite{Hlubina}. It has been also shown that, in the low-density limit at any fixed $U$, the ground state of the system is the $p$-wave BCS superfluid \cite{Chubukov}. The dynamic-cluster-approximation (DCA) simulations revealed (see \cite{Gull} and references therein)
a region of high-temperature  $d_{x^2-y^2}$-wave pairing  developing at
$U \gtrsim 6$. Nevertheless, the rich ground-state phase diagram guaranteed
by the above-mentioned findings remains elusive: So far, {\it none} of the phase boundaries is
known.

\begin{figure}[htbp]
\centering
\includegraphics[scale=1.0, angle=0, trim = 0 0 0 0, width=0.9\columnwidth]{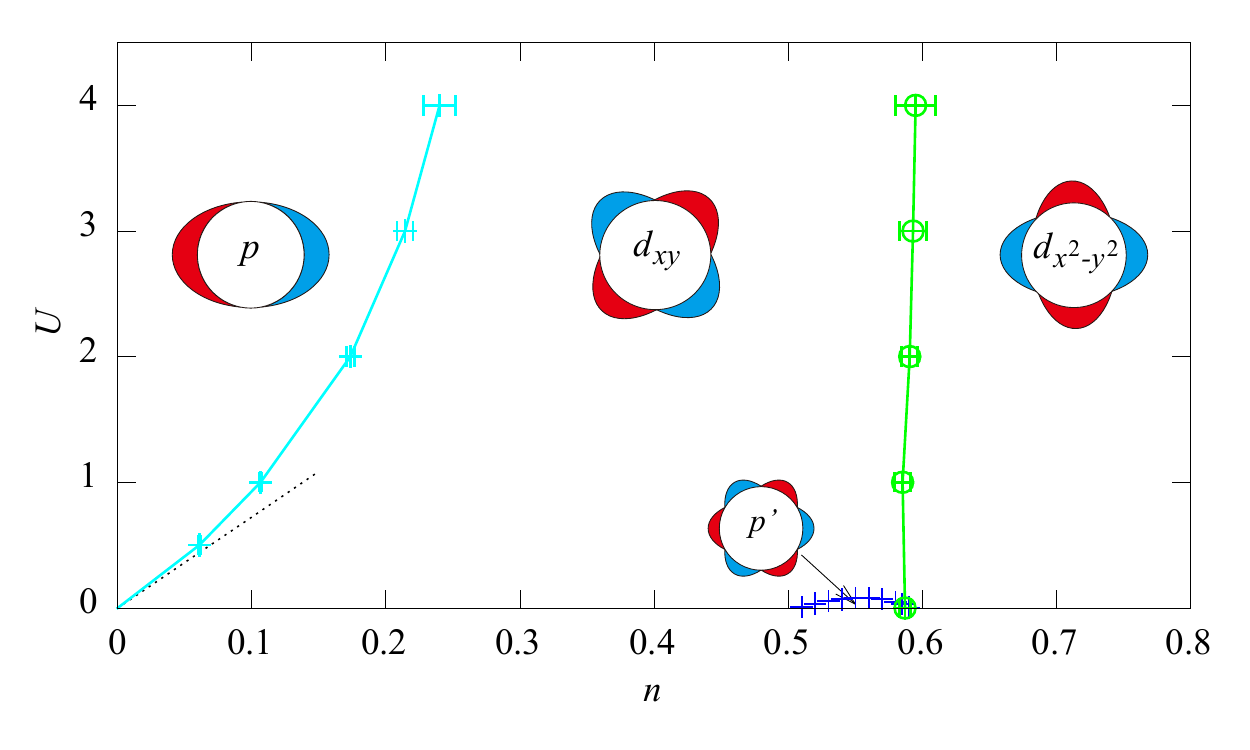}
\caption{Ground-state phase boundaries of the fermionic Hubbard model (\ref{FH}) in the emergent BCS regime. (Classification of superfluid phases in terms of the $D_{4h}$ group is explained in the text.)
The dashed straight line shows the $U \to 0$ limit } $n_c(U)=0.139 \, U$ for the $p$-$d_{xy}$ phase boundary. 
The $p'$ phase with six nodes exists only up to $U \approx 0.08$.
\label{fig:the_pd}
\end{figure}


{\bf Results.} ---We report accurate controllable results for a significant part of the ground-state phase diagram (Fig.~\ref{fig:the_pd}) of the Hubbard model, Eq.~(\ref{FH}), on the square lattice.
We concentrate on the region of moderate bare coupling $U \le 4$ and filling $n<0.7$, and first observe that there the system exhibits Landau Fermi-liquid behavior at temperatures $T_c < T \ll  E_F$; i.e., between the Fermi energy $E_F$ and the temperature of the superfluid phase transition $T_c \ll E_F$.
Hence we employ the first-principles theoretical framework consisting of: (i) asymptotically exact (in the $T_c/E_F \to 0$ limit) diagrammatic theory of Cooper instability in the Fermi liquid state~\cite{AGD,Gorkov} and (ii) unbiased Bold diagrammatic Monte Carlo (BDMC) simulation of the Fermi liquid parameters. We base our BDMC approach on the skeleton expansion in terms of the fully dressed interaction vertex in the particle-particle channel (analogous to the continuous-space technique developed in Refs.~\cite{BDMC_res_ferm,BDMC_res_ferm1} for the resonant Fermi gas) with an additional trick leading to near cancellation of large-amplitude contributions in the interaction vertex to improve numerical efficiency. In the considered regime of $U \le 4$, $n<0.7$,
the skeleton series is known to produce exact results~\cite{Kozik}, which we also checked explicitly by benchmarking the Green's function against the corresponding bare-series calculation. The approach allows us to controllably address all system's properties in the Landau Fermi-liquid regime
at temperatures $T_c < T \ll  E_F$ and deduce the leading channel for Cooper instability.

\begin{figure}[htbp]
\centering
\includegraphics[scale=1.0, angle=0, trim = 0 0 0 0, width=0.8\columnwidth]{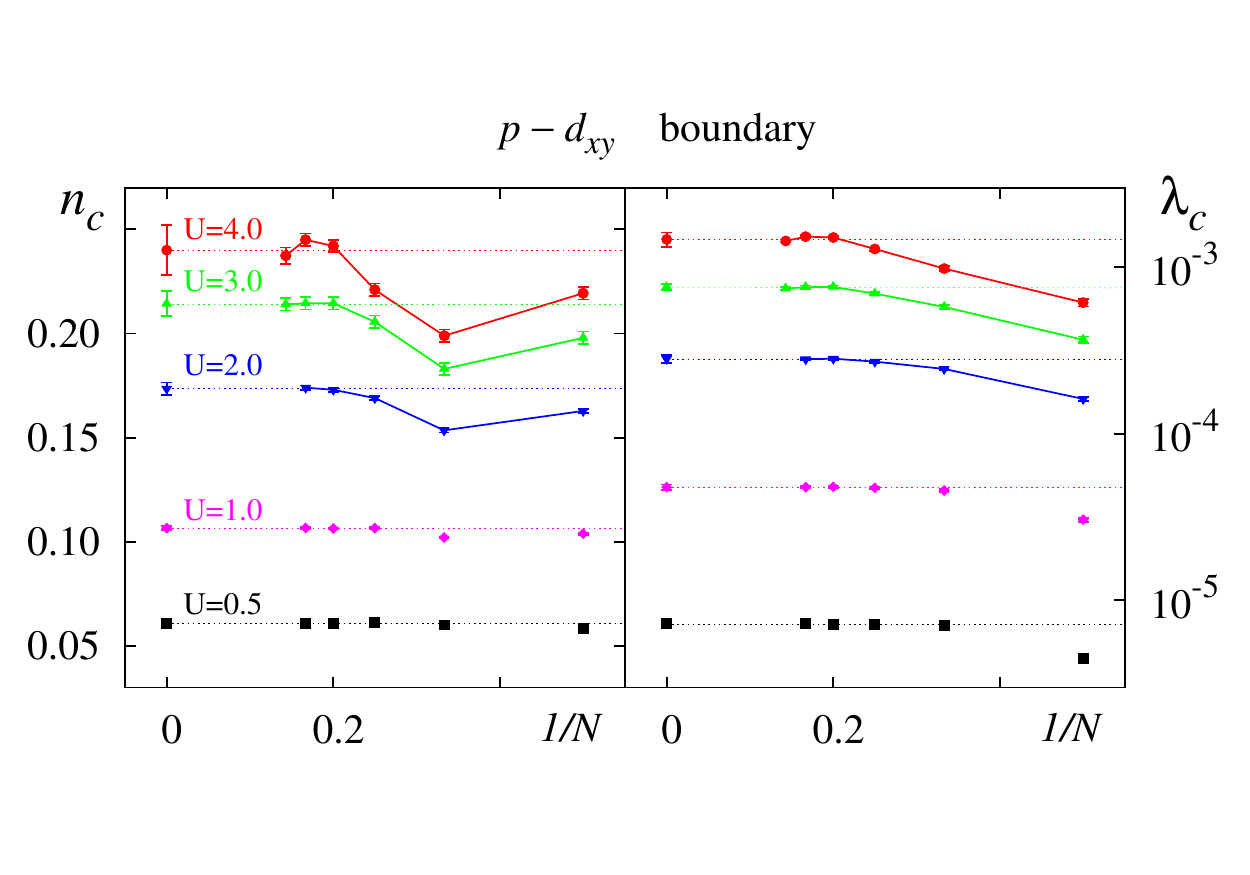} \\
\vspace{-5mm}
\includegraphics[scale=1.0, angle=0, trim = 0 0 0 0, width=0.8\columnwidth]{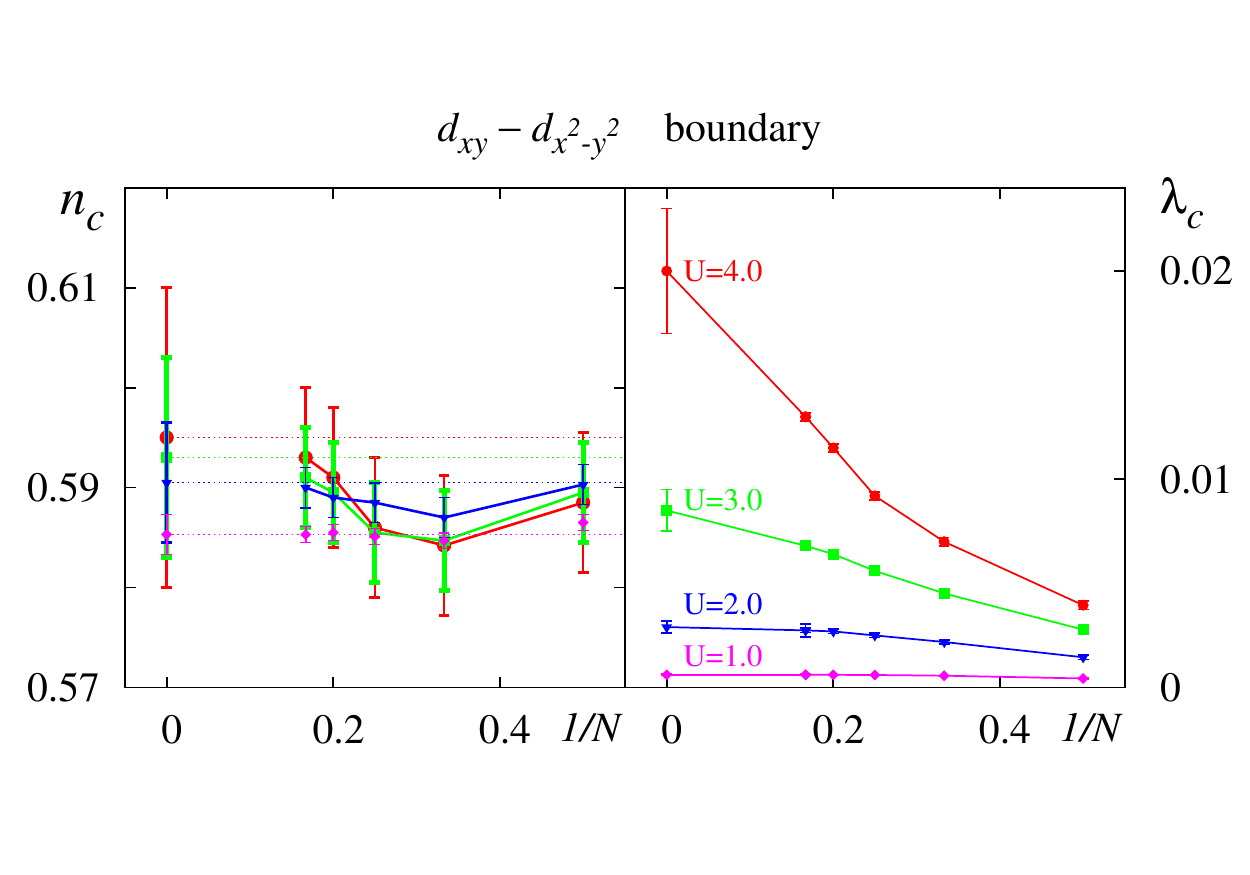}
\caption{Critical density and dimensionless coupling constants as functions of maximum diagram order, $N$, at various points along the phase boundaries (parameterized by $U$) with extrapolation to the $N\to \infty$ limit. Extrapolation was based on linear fits to the last three points. Error bars
were deduced from the stability of results when fits included four points.
}
\label{fig:g_of_N}
\end{figure}

Our main qualitative finding is that the effective (dimensionless) couplings in the Cooper channel remain small ($\le 0.1$) up to essentially non-perturbative values of the bare coupling $U \sim 4$ and densities up to $n <0.7$. This makes the problem of development of the Cooper instability amenable to controlled analytic treatment by diagrammatic perturbation theory. However, accurately determining the small effective coupling constants, shown in Fig.~\ref{fig:g_of_N}, up to $U=4$ requires a dramatic effort, involving development of essentially non-perturbative numeric techniques, such as a variant of BDMC employed here, and substantial computation time (see a discussion in the following section).

The revealed ground-state phase diagram is shown in Fig.~\ref{fig:the_pd}---where the error bars on the phase boundaries represent the full (systematic and statistical) error---and discussed in detail below.

{\bf Discussion.} ---Let us first focus on the lower left corner ($U \to 0$, $n \to 0$ limit) in Fig.~\ref{fig:the_pd}, which has been extensively studied by the perturbation theory 
in the $n \to 0$ and $U\to 0$ limit in Refs.~\cite{Baranov,Chubukov92, Chubukov}. Our $p$-$d_{xy}$ phase boundary is consistent with the linear law, $n_c = 0.139 U$, shown by the dashed line.  This behavior is
understood by comparing effective coupling constants derived in Refs.~\cite{Baranov,Chubukov92}
($\propto U^2$) and in Ref.~\cite{Chubukov} ($\propto U^3$). [The prediction for the slope from Refs.~\cite{Baranov,Chubukov92, Chubukov}
would be a factor of two smaller, $n_c/U = 0.069$. We attribute the discrepancy to typos:
Factors of two were missing in the density of states or/and the Cooper channel wave functions normalization.]

The phase diagram in the limit of $U \to 0$ for all densities $n$ has been obtained in second-order in $U$ calculations of Refs.~\cite{Hlubina, Raghu}. For $0.5< n <0.6$ the $p$-wave state is rather peculiar; we denote it $p'$ to emphasize the difference from the conventional $p$-wave, which was not fully addressed in Ref.~\cite{Hlubina}.
In a $p$-wave superfluid, the wavefunction of Cooper pairs has two nodes, in direct analogy with the case of the 
continuous rotation group (justifying the usage of the same symbol $p$).
In contrast, the pairing wavefunction of the $p'$-phase features six nodes \cite{remark2}.

Raghu \textit{et al.}~\cite{Raghu} generalized the second-order perturbation theory developed
in Ref.~\cite{Hlubina} to other types of lattices. However, in contrast to Ref.~\cite{Hlubina}, the $p'$ state at $0.5< n <0.6$ is apparently absent from their results for the square lattice. Apart from creating a controversy, the discrepancy circumstantially suggests that this part of the phase diagram might 
be very sensitive even at $U<1$.
This is indeed the case revealed here: the pocket of the $p'$ phase at $0.5< n <0.6$ vanishes already at $U \gtrsim 0.08$, as shown in Fig.~\ref{fig:the_pd}.
Our calculations also demonstrate that the BCS coupling constants $\lambda$ reported in Ref.~\cite{Raghu} are overestimated by a factor of $1/ \rho$, where $\rho$ is the Fermi-surface density of states, taking the values $1/4\pi \leq \rho \lesssim 0.185$ for $0 \leq n \lesssim 0.8$. This implies that from Ref.~\cite{Raghu} one can wrongly conclude that the superfluid $T_c$, which is exponentially sensitive to $\lambda$ ($T_c \sim E_F \exp(-1/\lambda)$) is orders of magnitude larger than it actually is, suggesting realization of high-temperature superconductivity by the Hubbard model already at very moderate $U$.

In general, our calculations reveal that perturbative $U \to 0$ results cannot be used to reasonably estimate the actual BCS coupling constants $\lambda$ (and thus the corresponding $T_c$).
Already for a weak interaction $U=1$,
the values of $\lambda$ at the $(d_{xy}-d_{x^2-y^2})$ boundary computed up to terms $\propto U^3$ are larger than those computed up to $U^2$ by a factor between two and three.

Even by state-or-the-art numeric techniques it still remains challenging to accurately and reliably compute basic physical quantities for the Hubbard model in the weak-to-intermediate coupling regime of $U \lesssim 4$.
For instance, for $(U=2, n=0.8, T=0.25)$ dynamical-cluster-approximation results
display significant oscillatory behavior as a function of the cluster size $L$ up to $L=98$ lattice sites, and thus accurate extrapolation to the thermodynamic limit $L \to \infty$ becomes difficult~\cite{Ferrero}.
Within the BDMC framework, one has to go substantially beyond the second-order \textit{skeleton} diagrams---convergence is observed only after accounting for diagrams of order $N=5$ (in terms of the fully dressed interaction vertex in the particle-particle channel and not the bare $U$) and above. Our scheme of combining BDMC and semi-analytic BCS treatment provides an effective and controllable method for studying correlated fermionic systems in the emergent BCS regime. The present calculation can be immediately generalized to other lattices or higher spatial dimensions, and, with minor modifications, can be used to explore other phase boundaries such as the antiferromagnetic transition   \cite{Otsuki,Schafer2015}.

Our most unexpected and essentially non-perturbative result here is the near-vertical boundary between different $d$-wave states at $n\approx 0.6$, see Fig.~\ref{fig:the_pd}. This behavior has nothing to do with the perturbation theory because (i) the boundary  {\it terminates} at the
$p'$ lobe at $U \approx 0.08$ (see Fig.~\ref{fig:the_pd}) implying that the $U \to 0$ theory fails at $U \approx 0.08$ already,
and (ii) the effective coupling $\lambda$ along the boundary has strong dependence on $U$ and the diagram order, see the lower 
right panel in Fig.~\ref{fig:g_of_N}.

Finally, we emphasize that in the $d_{x^2-y^2}$ phase, the dimensionless BCS
coupling $\lambda_{d_{x^2-y^2}}$ increases rapidly as the particle density $n$ and bare interaction $U$ are increased.
At the upper right corner of Fig.~\ref{fig:the_pd}, $(U=4, n=0.7)$, one already has $\lambda_{d_{x^2-y^2}} \sim 0.1$ corresponding to a critical temperature $T_c/E_F \sim 5 \times 10^{-5}$, typical for conventional superconductors. This is still well within the domain of the emergent BCS regime, but
if $\lambda_{d_{x^2-y^2}}$ increases further by $50 \% $ at larger $U$ and $n$, the system would
meet the criterion for being a high-temperature superconductor.
As Fig.~\ref{fig:g_of_N} suggests, the value of $\lambda$ may indeed be significantly
enhanced as $U$ is further increased from $U=4$.
Unfortunately, the present BDMC formulation fails in such strongly correlated regime~\cite{Kozik},
and novel techniques/approaches dealing with diagrammatic series need to be developed \cite{Toschi,Rubtsov}. Nevertheless, even within the present approach, by computing $T_c(n,U)$ (with correct pre-exponential factor \cite{Phillips,in_progress}) up to the $(U=4, n=0.7)$ corner, one can obtain a reasonable  extrapolation to the intriguing region of $(U \approx 6, n\approx 0.85)$ thus shedding a significant light on the crossover from emergent  BCS to  high-$T_c$ regime expected to take place in this range of parameters \cite{Gull}.

{\bf Methods.} ---The {\it emergent BCS regime} is a Cooper instability due to weak effective attraction between the quasiparticles developing
in a strongly correlated fermionic system at energy scales much smaller than the Fermi energy.
In this scenario, as the temperature is decreased, the system first enters the standard Fermi liquid (FL) state characterized by renormalized quasiparticle properties and effective interactions. The quasiparticle Green's function in the vicinity of the Fermi surface [$|\xi| \ll E_F $ and $| k  - {\kern 1pt} {k_F(\hat{k})} | \ll k_F(\hat{k})$] takes on the form:
\begin{equation}
G({\bf{k}},\xi ) \, \approx \, \frac{z(\hat{k})}{i\xi  - {\kern 1pt} {{\bf{v}}_F}(\hat{k}) \! \cdot \! [{\bf{k}}  - {\kern 1pt} {{\bf k}_F(\hat{k})}] }  \; ,
\label{G_FL}
\end{equation}
where ${\bf k}$ is the momentum and $\xi$ is the Matsubara frequency, and the Fermi surface is parameterized in terms of the Fermi momentum ${\bf k}_F(\hat{k})$ in the direction of $\hat{k}$, with ${\bf v}_F(\hat{k})$ and $z(\hat{k})$ being the Fermi velocity and quasiparticle residue respectively. The Cooper instability then develops
logarithmically slowly in the FL state and is marked by divergence of pairing susceptibility at the transition
temperature $T_c$ that is exponentially small compared to the FL energy scale.

\begin{figure}[htbp]
\centering
\includegraphics[scale=1.0, angle=0, trim = 0 13cm 0 0, width=0.9\columnwidth]{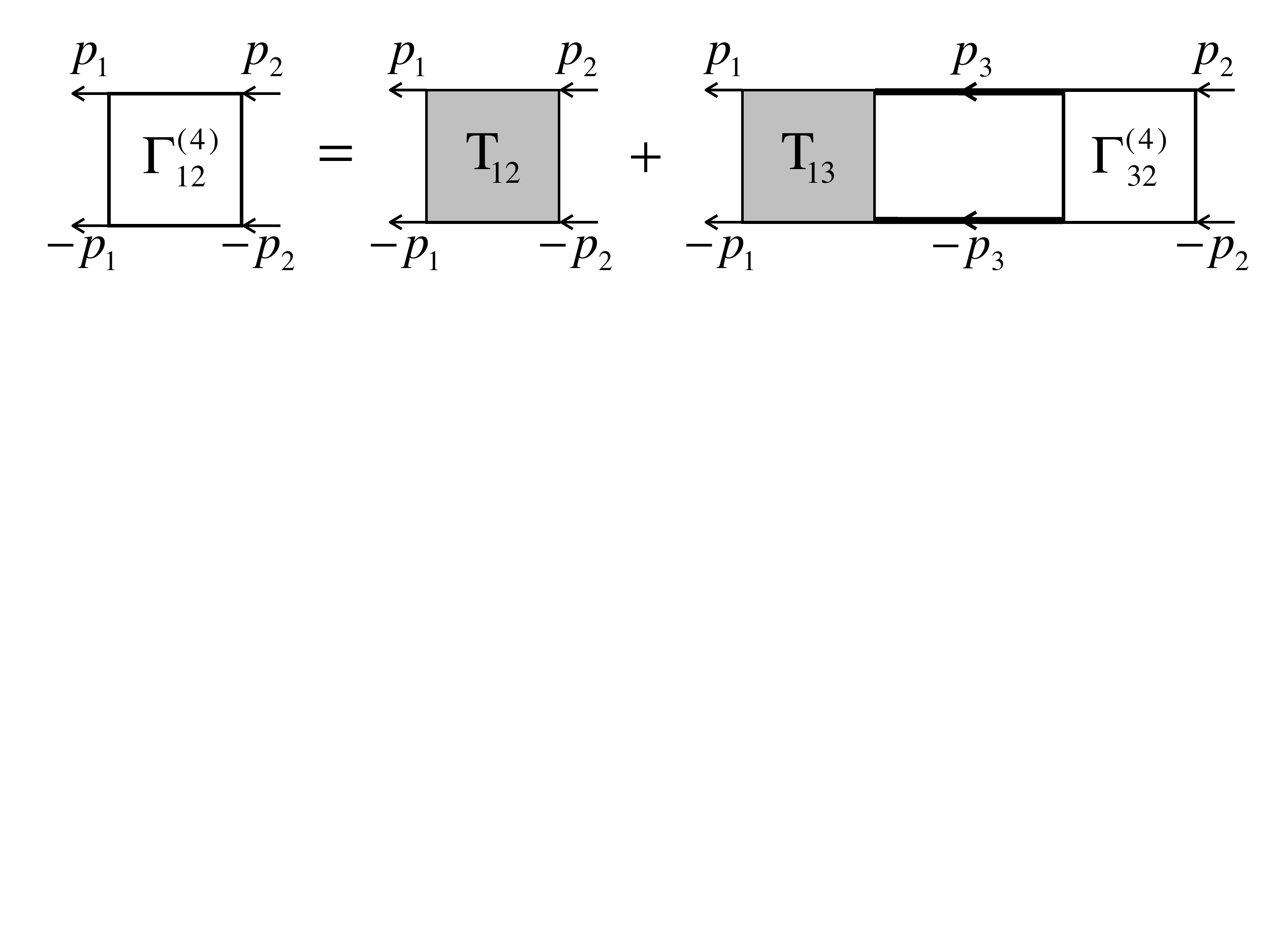}
\caption{ Bethe-Salpeter equation for $\Gamma^{(4)}$, where $p_i \equiv (\xi_i, \mathbf{k}_i)$.
}
\label{fig:Cooperon}
\end{figure}

Physically, this behavior is typical for models with local repulsive coupling, where
weak attractive effective interactions---described by the irreducible (in the particle-particle channel)
four-pole vertex ${\rm T}$---are an emergent low-energy property. By definition, ${\rm T}$ is the sum of
all four-pole diagrams that can not be split into disconnected pieces by cutting two particle lines.
From the Bethe-Salpeter relation, Fig.~\ref{fig:Cooperon}, for the full four-pole vertex $\Gamma^{(4)}$,
we see that the smallness of the attractive part of ${\rm T}$ is a natural condition preventing
$\Gamma^{(4)}$ from dramatic growth at $T\ll E_F$. Indeed, in the FL state, the leading contribution to the
integral over ${\bf k}_3$ in the second term in the r.h.s. of Fig.~\ref{fig:Cooperon} comes from
$\int d^d k_3 \sum_{\xi_3}G(p_3)G(-p_3)$ in close vicinity to the Fermi surface,
where only the finite temperature (i.e., discreteness of Matsubara frequency $\xi_3$) prevents it
from logarithmic divergence. With the logarithmic accuracy at $T\ll E_F$, we have
\begin{equation}
\Gamma^{(4)}_{\hat{k}_1,\hat{k}_2}   \,  \approx \,  {\rm T}_{\hat{k}_1,\hat{k}_2} + \ln  {E_F \over T} \int\! {\rm T}_{\hat{k}_1,\hat{k}_3} Q_{\hat{k}_3}\Gamma^{(4)}_{\hat{k}_3,\hat{k}_2} d^{d -\!1}  \hat{k}_3   ,
\label{B-S}
\end{equation}
where $\Gamma^{(4)}_{\hat{k}_1,\hat{k}_2}$ and ${\rm T}_{\hat{k}_1,\hat{k}_2}$ are $\Gamma^{(4)}$ and ${\rm T}$
at vanishing frequencies projected to the Fermi surface:
\begin{eqnarray}
\Gamma^{(4)}_{\hat{k}_1,\hat{k}_2}\equiv \Gamma^{(4)}( {\bf k}_1\! = \! {\bf k}_F(\hat{k}_1),\xi_1\to 0;   {\bf k}_2\! = \! {\bf k}_F(\hat{k}_2),\xi_2\to 0) \, ,\nonumber
\end{eqnarray}
and $Q_{\hat{k}}$ is the product of $z^2(\hat{k})$ and the single-component density of states at the $\hat{k}$-point
on the Fermi surface. The systematic error in (\ref{B-S}) comes from the ultra-violet cutoff scale,
$E_F \to c E_F$, where $c$ is some order-unity factor~\cite{remark1}.

Switching to the matrix notations, $\Gamma^{(4)}_{\hat{k}_1,\hat{k}_2} \! \to\! \hat{\Gamma}^{(4)}$, ${\rm T}_{\hat{k}_1,\hat{k}_2} \! \to \! \hat{\rm T}$, ${\rm T}_{\hat{k}_1,\hat{k}_2}Q_{\hat{k}_2}\! \to\! \hat{\rm M}$,
we find
\begin{equation}
\hat{\Gamma}^{(4)} \,  \approx \,  \left[ 1-\ln (E_F/T) \hat{\rm M}\, \right]^{-1} \hat{\rm T}\,  ,
\label{matrixwise}
\end{equation}
implying that $\Gamma^{(4)}$---and thus the static response function in the Cooper channel---diverges at the critical temperature
\begin{equation}
T_c \, =\, c\, E_F\, e^{-1/\lambda } \, ,
\label{T_c}
\end{equation}
where $\lambda$ is the largest positive eigenvalue of $\hat{\rm M}$. The consistency
of the emergent BCS picture based on weak Cooper instability requires $\lambda \ll 1$.

Solving the problem with logarithmic accuracy amounts then to finding the eigenvalues/eigenvectors
of a real symmetric matrix
\begin{equation}
{\cal T}_{\hat{k},\hat{k}'}  \psi_{\hat{k}'}  = \lambda \psi_{\hat{k}}\,  , \qquad  {\cal T}_{\hat{k},\hat{k}'} = Q_{\hat{k}}^{1\over 2} {\rm T}_{\hat{k},\hat{k}'} Q_{\hat{k}'}^{1\over 2}\, ,
\label{eigenvalue_problem}
\end{equation}
where the eigenvector $\psi_{\hat{k}}$ is the wave function of the Cooper pair in the momentum representation.

{\it Parameterization and $D_{4h}$ nomenclature in 2D.} In two dimensions, it is convenient to parameterize $\hat{k}$ with the polar angle $\theta$,
and to write the eigenvalue/eigenvector problem explicitly as
\begin{eqnarray}
 \int_0^{2\pi} \!  {\cal T}_{\theta,\theta'} \psi_{\theta'} {d \theta'\over 2\pi} &=& \lambda \psi_\theta\, , \qquad
 {\cal T}_{\theta,\theta'} = Q_{\theta}^{1\over 2} {\rm T}_{\theta,\theta'} Q_{\theta'}^{1\over 2} \, ,  ~~
\label{eigenvalue_problem_2D} \\
Q_{\theta}\,  &=&\,
k_F(\theta)\, z^2 (\theta)/[2\pi\,  \hat{\theta}\! \cdot\! {\bf v}_F(\theta)]  \; .
\label{Q}
\end{eqnarray}
By the $D_{4h}$ symmetry of the square lattice, $ {\cal T}_{\theta,\theta'}$ splits into five
independent blocks corresponding to $s$, $p$, $d_{x^2-y^2}$, $d_{xy}$, and $g$ eigenvector sectors.
The $p$-sector is doubly degenerate and can be further split into two independent sectors, $p_x$ and $p_y$, related to each other by  $\pm \pi/2$ rotations. For each of the six (sub)sectors, the symmetry properties
of the corresponding vectors $f(\theta)$ are readily seen from their Fourier expansions
($m$ is integer):
\begin{eqnarray}
\small
f_{s}(\theta) &=& \! \sum_{m=0}^{\infty} A_m\cos(4m\theta) , \nonumber \\
f_{g}(\theta) &=& \! \sum_{m=1}^{\infty} B_m \sin(4m\theta)  ,\nonumber \\
f_{\scriptsize \!\! \left\{ \!\!\!{\begin{array}{*{20}{c}}{p_y}\\{p_x}\end{array}}\!\!\!\right\} }\! \! (\theta) \, &=&   \sum_{m=0}^{\infty} \, C_m
 \left\{ {\begin{array}{*{20}{c}} {\cos\,  [(2m+1)\theta] }\\ {\sin \, [(2m+1)\theta]  } \end{array}} \right\} ,  \label{p_wave} \\
f_{\scriptsize \!\! \left\{ \!\!\!{\begin{array}{*{20}{c}}{d_{x^2\! -y^2}}\\{d_{xy}}\end{array}}\!\!\!\right\} }\! \! (\theta)
 \, &=& \sum_{m = 0}^\infty  \,
  \left\{ {\begin{array}{*{20}{c}} { D_m \cos\, [(4m+2)\theta]}\\ {  E_m \sin \, [(4m+2)\theta] } \end{array}} \right\} . \nonumber
\end{eqnarray}
The  $f_{s}$ is invariant with respect to all point-group operations;
$f_{g}$ is invariant with respect to $\pi/2$ rotations, but changes its sign under each of the
four $D_{4h}$ reflections; $f_{p_y}$/$f_{p_x}$ is symmetric with respect to reflections over the $x$/$y$-axis and
anti-symmetric with respect to reflections over the $y$/$x$-axis (also, the $\pi/2$ rotation of $f_{p_y}$ turns it into $f_{p_x}$).
The functions in both $d$ sectors change their sign when rotated by $\pi/2$:
$f_{d_{x^2\! -y^2}}$/$f_{d_{xy}}$ is symmetric/anti-symmetric with respect to reflections
over $x$ and $y$ axes, and anti-symmetric/symmetric with respect to reflections by $\pm\pi/4$ axes.
Fermionic anti-symmetry implies a spin-triplet state for $p$-wave pairing and a spin-singlet state for the other four sectors.

\begin{figure}[htbp]
\centering
\includegraphics[scale=1.0, angle=0, trim = 0 14cm 0 0, width=0.9\columnwidth]{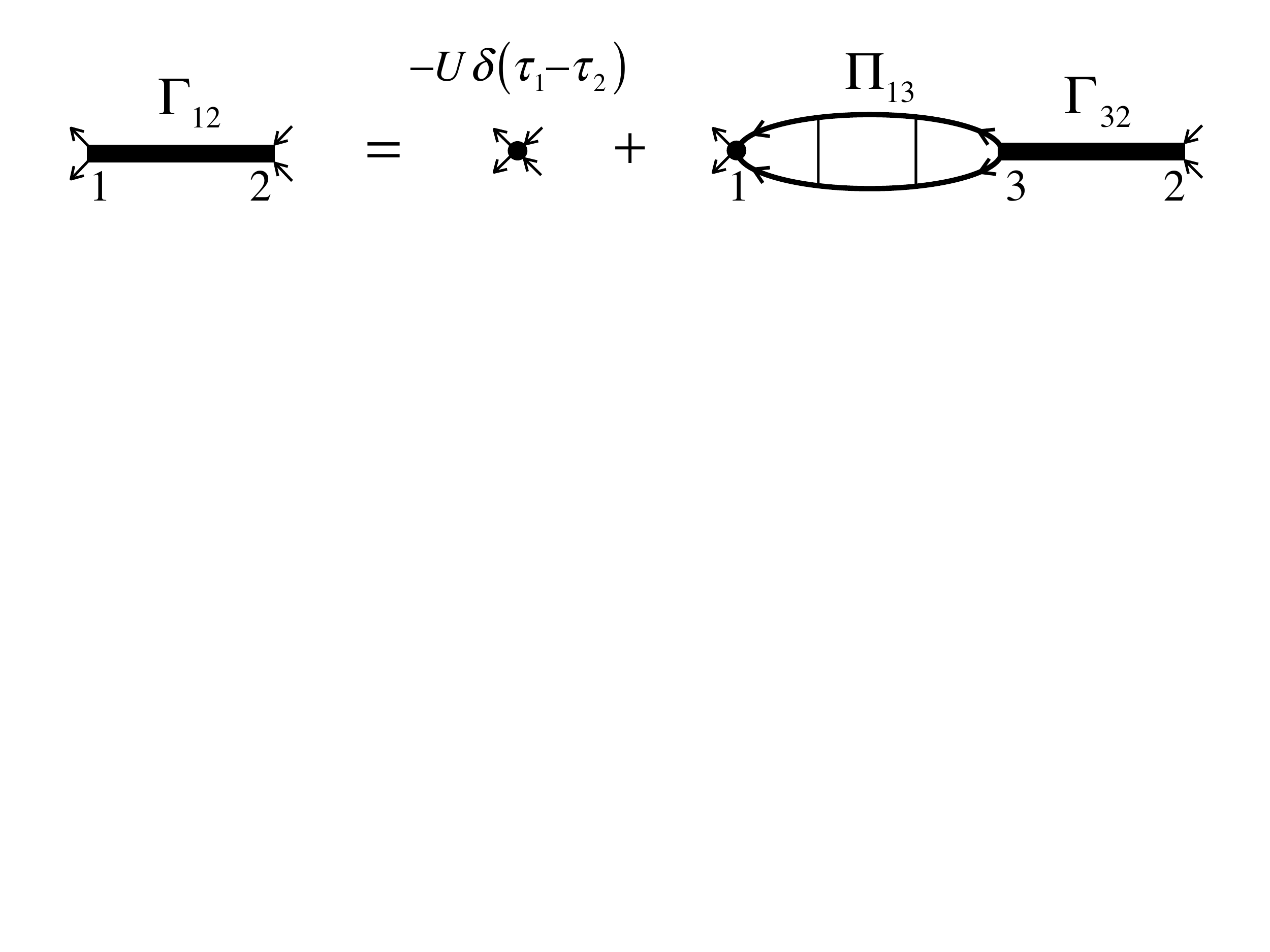}
\caption{ The $\Gamma$-line in the time-momentum representation
(spin and momentum indexes are suppressed for clarity):
$\Gamma_{12}\equiv \Gamma(\tau_1-\tau_2)$, $\Pi_{13}\equiv \Pi(\tau_1-\tau_3)$, etc.
Integration over internal times is assumed. Pair self-energy $\Pi_{13}$ is the sum of
all vertex-irreducible diagrams starting, at time $\tau_3$, and ending, at time $\tau_1$,
with spin-up and spin-down outgoing (incoming) single particle propagators.
A diagram is vertex-irreducible if it remains connected after cutting across any single interaction vertex.
The lowest-order diagram contributing to $\Pi_{13}$ is a pair of dressed propagators going
from $\tau_3$ to $\tau_1$. The second term in the r.h.s. is a continuous function
of $\tau$ and will be referred to as $\tilde{\Gamma}$. Hence, $\Gamma(\tau, {\bf k})=-U\delta(\tau) + \tilde{\Gamma}(\tau, {\bf k})$.}
\label{fig:gamma}
\end{figure}

{\it BDMC method with the ladder-summation trick.} Similar to the system of resonant fermions, the
locality of interaction allows one to introduce propagators based on interaction vertexes and
pairs of fermions (and to fully dress them) by considering sums of ladder diagrams,
see Ref.~\cite{BDMC_res_ferm} and Fig.~\ref{fig:gamma}. Effectively, this amounts to replacing
the bare interactions in Feynman diagrams with exact two-body scattering amplitudes; this trick
is particularly important for dealing with strong interaction in the dilute gas limit
by eliminating the expansion in a large parameter.

There is, however, a technical difficulty in combining bare interaction with ladder terms
in the imaginary time representation: The first term in the r.h.s. of Fig.~\ref{fig:gamma}
is a generalized $-U\delta (\tau)$ function (for resonant fermions, this term vanishes upon
taking the zero-range limit), while the rest of the diagrams, $\tilde{\Gamma}$, is a continuous
function of $\tau$. The effective smallness of $\Gamma $ in the dilute-gas regime at large $|U|$
takes place only under $\tau$-integration and mathematically happens as follows. For large but finite
$U$ the sum of ladder diagrams behaves as a regularized $U\delta (\tau)$ function; i.e.,
the range of variation of $\tilde{\Gamma}(\tau)$ is $\sim 1/U$ while its amplitude
is such that $\int_0^{\tau_0}\tilde{\Gamma}(\tau)\, d\tau \approx U $  for
$\tau_0 \gg 1/|U|$.  In Monte Carlo methods, however, the integration is achieved by sampling the
{\it integrands} with the weighting factors proportional to their absolute values, meaning that
a na\"{i}ve scheme will sample $- U\,\delta (\tau)$ terms separately from $\tilde{\Gamma}(\tau)$
terms and their mutual compensation will be revealed only in the painful statistical limit.

The solution is to transform the functional form of the bare vertex to make it
(i)  compatible with that of $\tilde{\Gamma}(\tau)$ at the level of \textit{integrands}, and
(ii) such that the diagram value remains intact under integration.
To this end we introduce a function $\tilde{\Gamma}_U$ with the following properties
\begin{equation}
\int_0^{\beta}\tilde{\Gamma}_U( \tau) \, d \tau = -U \, . 
\label{Gamma_U}
\end{equation}
The particular design of $\tilde{\Gamma}_U(\tau)$ still has a freedom.
We choose $\tilde{\Gamma}_U(\tau) =  -\tilde{\Gamma}(\tau) +c_0$, where  $c_0$ is a constant of order unity or much smaller.
This guarantees that, for  $|U|\gg 1$, the condition of compensation, $\tilde{\Gamma}_U(\tau) \approx  -\tilde{\Gamma}(\tau)$, is satisfied.

\begin{figure}[htbp]
\centering
\includegraphics[scale=1.0, angle=0, trim = 0 10cm 0 0, width=0.9\columnwidth]{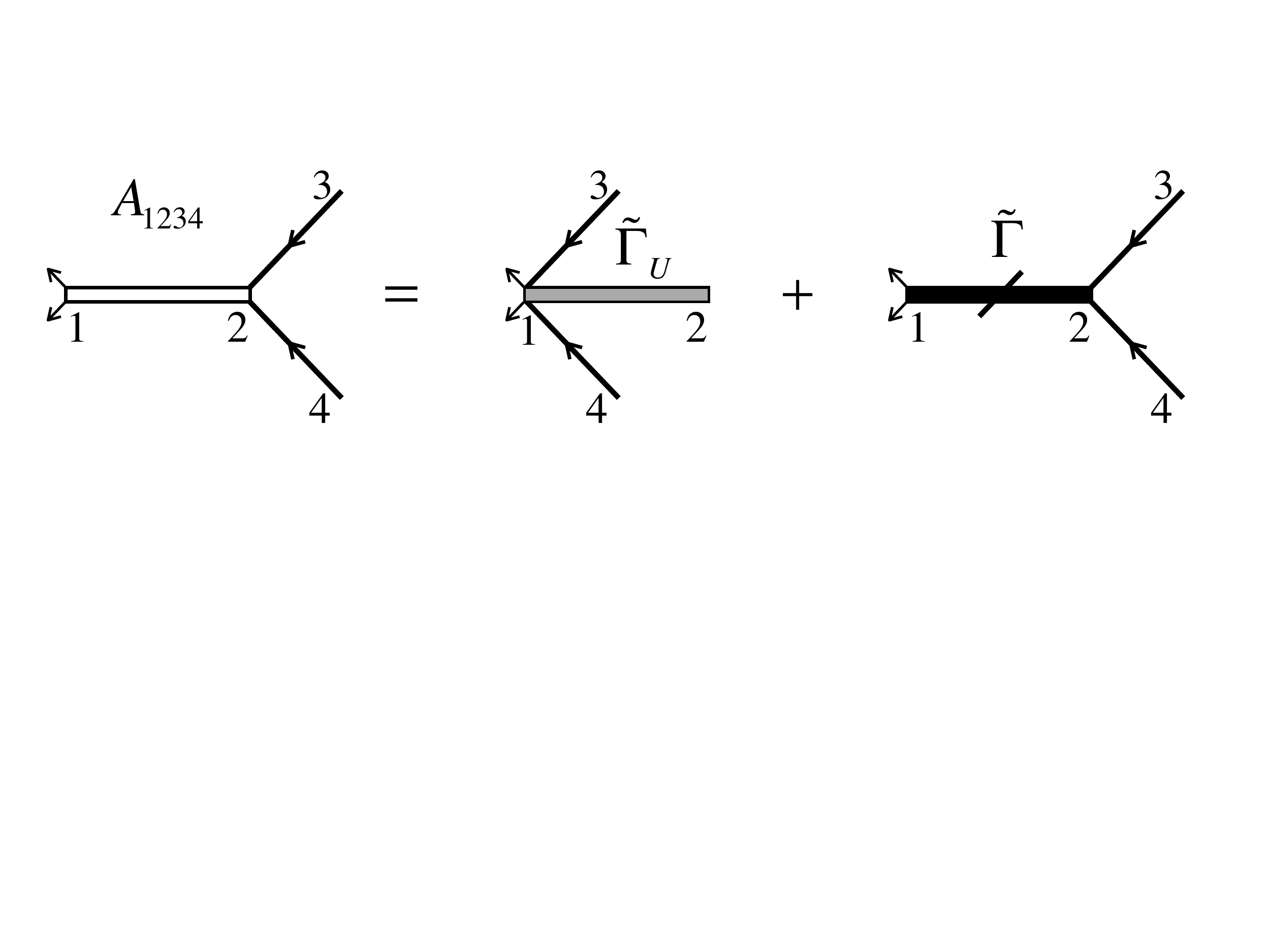}
\caption{The compensation trick. The diagram element $A_{1234}$
is understood as the sum of two terms with different assignment of the end point
for incoming fermionic propagators.}
\label{fig:gamma_new}
\end{figure}

We then formally---and identically---represent each coupling constant $U$ in the diagrammatic series as an integral over the auxiliary time variable associated with the bare vertex,
thereby replacing the bare vertex with the $\tilde{\Gamma}_U$ function as pictured graphically in Fig.~\ref{fig:gamma_new}. Since $\tilde{\Gamma}_U$ has the same functional
structure as $\tilde{\Gamma}$, we sum up the two elementary diagrammatic contributions into one,
$A_{1234}$, as shown in Fig.~\ref{fig:gamma_new}. Thereby, we arrive at the diagrammatic formulation
identical to that for resonant fermions \cite{BDMC_res_ferm}, but with a modified rule for reading
the diagram value: The single diagram element $A_{1234}$, now contributes a factor
(momenta are suppressed for clarity)
\begin{eqnarray}
A_{1234}\, &=& \, {{\tilde \Gamma }_U}({\tau _1} - {\kern 1pt} {\tau _2})\,{G_ \uparrow }({\tau _1} - {\kern 1pt} {\tau _3})\,{G_ \downarrow }({\tau _1} - {\tau _4}) \nonumber
\\&+& \, \tilde \Gamma ({\tau _1} - {\kern 1pt} {\tau _2})\,{G_ \uparrow }({\tau _2} - {\kern 1pt} {\tau _3})\,{G_ \downarrow }({\tau _2} - {\tau _4}) \;,
\label{A}
\end{eqnarray}
to the integrand of the diagram it enters. It contains two terms that become close in absolute values
and opposite in sign when $|{\tau _1} - {\kern 1pt} {\tau _2}|\lesssim 1/|U|$. This is how the
large-$U$ compensation is achieved at the level of integrands.
Apart from this specific way of evaluating the diagram value, the rest of the BDMC protocol is
essentially identical to that for resonant fermions~\cite{BDMC_res_ferm}.

{\it Numeric analysis.} We employ the following protocol dictated by FL physics. We start with the BDMC simulation of the single-particle Green's function at some temperature $T\ll E_F$, low enough for observing a sharp Fermi-step in the momentum distribution,
and extract all quasiparticle FL parameters.
We then use this Green's function to perform the BDMC simulation
of the irreducible vertex $ {\rm T}_{\hat{k}_1,\hat{k}_2}$, extract eigenvalues/eigenfunctions for all Cooper channels by solving the eigenvalue problem (\ref{eigenvalue_problem_2D}), and locate the phase boundaries
from points where $\lambda$ for the two competing ground-state phases coincide. All simulations are performed with explicit truncation of diagrammatic series at some maximum order $N$. Extrapolation with respect to $N$ brings the corresponding systematic error under control, see Fig.~\ref{fig:g_of_N}.

We repeat these simulations at different temperatures to ensure that final results are temperature-independent. 
To eliminate the slowly vanishing (and quite substantial in the $n \to 0$ limit) finite-temperature correction
to $ {\rm T}_{\hat{k}_1,\hat{k}_2}$, we observe that the leading term in
this correction comes from the second-order diagram and hence calculate the
corresponding contribution (semi-analytically) directly at $T=0$. In addition, we have verified that spin and density correlations
(particle-hole channels) do not exhibit any flow towards instability at low temperature for $(n=0.6,~U=4)$.


{\bf Conclusion.} ---Within the diagrammatic framework based on the unbiased BDMC method and asymptotically exact (in the weak-effective-coupling limit) theory of Cooper instability in the Fermi-liquid state, we  revealed a significant part of rich ground-state phase diagram of the fermionic Hubbard model on the square lattice. Specifically, we addressed the region of moderate bare coupling $U/t \le 4$ and filling $n<0.7$, where the system was found to exhibit Landau Fermi-liquid behavior within a broad temperature interval between the Fermi energy  $E_F$ and the superfluid transition temperature $T_c  \ll  E_F$---a signature
of the (emergent) weak effective coupling in the Cooper channel.  The main reason why our data are confined to $U \le 4$ and $n<0.7$ is the convergence of diagrammatic series; it becomes problematic outside this range of parameters.

\acknowledgments
We are grateful to Kris Van Houcke and and F\'elix Werner for sharing their expertise in
code development, and to A. Chubukov, M. Baranov, E. Gull, J. Gukelberger, and M. Troyer for valuable discussions. We thank X.-W. Liu for his participation in the study of convergence of skeleton series. This work was supported by the Simons Collaboration on the Many Electron Problem,
National Science Foundation under the grant PHY-1314735, the MURI Program ``New Quantum Phases of Matter" from AFOSR, and the Swiss National Science Foundation,
NSFC Grant No. 11275185, CAS, and NKBRSFC Grant No. 2011CB921300.
We acknowledge the hospitality of Kavli Institute for Theoretical Physics China at Beijing. 



\begin{thebibliography}{99}
\bibitem{remark1} Methods for systematic (in parameter $g\ll 1$) evaluation of $c$ are well-known \cite{Gorkov}
                  and will be discussed elsewhere.
\bibitem{remark2}  In terms of the expansion (\ref{p_wave}), the nodal structure of the $p$-wave state is dominated by $\cos[\theta]$, while the nodal structure of the $p'$-eigenvector
    comes from $\cos[3\theta]$, rendering it akin to the $f$-wave state in continuous space.
\end{thebibliography}
\end{document}